\documentclass[review]{elsarticle}

\usepackage{lineno,hyperref}
\modulolinenumbers[5]

\usepackage{graphicx}
\usepackage{bm}
\hypersetup{colorlinks=false}

\journal{Annals of Physics}

\begin{document}

\begin{frontmatter}

\title{Exact solutions of the Klein-Gordon equation in the Kerr-Newman background and Hawking radiation}


\author[mymainaddress]{H. S. Vieira\corref{mycorrespondingauthor}}
\cortext[mycorrespondingauthor]{Corresponding author}
\ead{horacio.santana.vieira@hotmail.com}

\author[mymainaddress]{V. B. Bezerra}
\ead{valdir@fisica.ufpb.br}

\author[mysecondaryaddress]{C. R. Muniz}
\ead{celiomuniz@yahoo.com}

\address[mymainaddress]{Departamento de F\'{i}sica, Universidade Federal da Para\'{i}ba, Caixa Postal 5008, CEP 58051-970, Jo\~{a}o Pessoa, PB, Brazil}
\address[mysecondaryaddress]{Grupo de F\'{i}sica Te\'{o}rica (GFT), Universidade Estadual do Cear\'{a}, UECE-FECLI, Iguatu, CE, Brazil}

\begin{abstract}
This work considers the influence of the gravitational field produced by a charged and rotating black hole (Kerr-Newman spacetime) on a charged massive scalar field. We obtain exact solutions of both angular and radial parts of the Klein-Gordon equation in this spacetime, which are given in terms of the confluent Heun functions. From the radial solution, we obtain the exact wave solutions near the exterior horizon of the black hole, and discuss the Hawking radiation of charged massive scalar particles.
\end{abstract}

\begin{keyword}
charged massive scalar field \sep confluent Heun function \sep black hole radiation
\MSC[2010] 81Q05 \sep 83C45 \sep 83C57 \sep 83C75
\end{keyword}


\end{frontmatter}

\linenumbers
%
%
\section{Introduction}
The study of the interaction of quantum systems with gravitational fields goes back to the beginning of the last century, when the generalization of quantum mechanics to curved spaces was discussed, motivated by the idea of constructing a theory combining quantum physics and general relativity \cite{AnnPhys.386.109,ZPhys.37.895,ZPhys.40.117,ZPhys.41.407,ProcRSocLondA.117.610,ProcRSocLondA.118.351,ZPhys.53.592,ZPhys.57.261,AnnPhys.410.305,AnnPhys.410.337,Physica.6.899}. Along this line of research the investigation of solutions of the Klein-Gordon equation in some gravitational fields as well as their consequences has been discussed in the literature \cite{rowan2,stephenson,JPhysAMathGen.10.15,JMathPhys.22.1457,JMathPhys.26.2286,pimentel,elizalde,pimentel2,semiz,kraori,JMathPhys.40.4538,ProgTheorPhys.112.983,bezerra,vakili,sandro,GenRelativGrav.43.833}. It is worth emphasizing that the study of the behavior of scalar fields in black hole backgrounds could be used, in principle, to understand the physics of these objects. Therefore, it is important to find solutions of the Klein-Gordon equation for real as well for complex fields and analyze the phenomena related to them, as for example, the radiation of scalar particles.

In a recent paper \cite{ClassQuantumGrav.31.045003}, we obtained the exact solutions of the Klein-Gordon equation for a massive real scalar field in the Kerr-Newman spacetime, valid in the whole space that corresponds to the black hole exterior, which means between the exterior event horizon and infinity. They are given in terms of the confluent Heun functions \cite{Ronveaux:1995,Slavyanov:2000}. Thus, we extended the range in which the solutions are valid as compared with the ones obtained by Rowan and Stephenson \cite{JPhysAMathGen.10.15}, which is valid near the exterior horizon and at infinity.

Hawking radiation \cite{CommunMathPhys.43.199} is an interesting phenomenon concerning the emission of any types of particles by black holes. In order to study this phenomenon, many different methods has been proposed \cite{PhysRevD.11.1404,PhysRevD.13.2188,PhysRevD.14.332,PhysRevD.15.2738,ProcRSocLondA.15.2738,GenRelativGravit.20.239,PhysRevLett.85.5042,PhysLettB.642.124,IntJModPhysA.22.1705}. In particular, the emission of scalar particles has also been discussed in the literature \cite{PhysRevD.82.044013,PhysLettB.692.61,PhysLettB.697.398}.

Recently, Zhang and Zhao \cite{PhysLettB.618.14} studied the Hawking radiation of a Kerr-Newman black hole by introducing a very particular coordinate system. The coordinates have some attractive properties: the time direction is a Killing vector, the metric is smooth at the horizons, among others. Huaifan \textit{et al}. \cite{ChinJPhys.47.618} introduced the tortoise coordinates and extended the classical Damour-Ruffini method \cite{PhysRevD.14.332} to discuss the general radiation spectrum. Umetsu \cite{IntJModPhysA.25.4123} studied the Hawking radiation by using the tunneling mechanism process and the dimensional reduction near the horizon.

In our present paper we obtain the exact solutions of the Klein-Gordon equation for a charged massive scalar field in the background under consideration, valid in the whole spacetime that corresponds to the black hole exterior, which means between the exterior event horizon and infinity. Using the radial solution which is given in terms of the confluent Heun functions and taking into account their properties, we study the Hawking radiation of charged massive scalar particles.

A similar result concerning massless scalar particles in the Kerr-Newman back\-ground is already in the literature \cite{ProgTheorPhys.100.491,ProgTheorPhys.102.253,JMathPhys.48.042502,ClassQuantumGrav.27.135001}. These were obtained using the Teukolsky equation \cite{AstrophysicalJ.185.635} which describes the dynamical scalar perturbations, as well as gravitational, electromagnetic and spinor, of a rotating black hole. For the scalar perturbation, the solutions of the Teukolsky equation, which can be mapped into the Klein-Gordon equation for a massless field, are also given in terms of the confluent Heun functions.

This paper is organized as follows. In Section 2, we present some features of the Kerr-Newman spacetime and some elements relevant to study the Hawking radiation. In Section 3, we introduce the Klein-Gordon equation for a charged massive scalar field in a curved background and write down this equation in the Kerr-Newman spacetime, separate the angular and radial parts, and we present the exact solution of both angular and radial equations. In section 4, we obtain the solutions for ingoing and outgoing waves, near to the exterior horizon of the black hole. In section 5, we extend the wave solutions from the outside of the black hole into the inside of the black hole. In Section 6, under the condition that the spacetime total energy, total charge and total angular momentum are conserved, we derive the black hole radiation spectra taking into account the reaction of the radiation to the spacetime. Finally, in section 7 we present our conclusions.
%
%
\section{Kerr-Newman spacetime}
The metric generated by a black hole with angular momentum per unit mass $a=J/M$, electric charge $Q$, and mass (energy) $M$ is the Kerr-Newman metric \cite{MTW:1973}, whose line element, in the Boyer-Lindquist coordinates \cite{JMathPhys.8.265}, is given by
\begin{eqnarray}
ds^{2} & = & \frac{\Delta}{\rho^{2}}\left(dt-a\sin^{2}\theta\ d\phi\right)^{2}-\frac{\rho^2}{\Delta}dr^{2}-\rho^{2}\ d\theta^{2}\nonumber\\
& - & \frac{\sin^2\theta}{\rho^2}\left[\left(r^2+a^2\right)d\phi-a\ dt\right]^{2}\ ,
\label{eq:metrica_Kerr-Newman}
\end{eqnarray}
where
\begin{equation}
\Delta=r^{2}-2Mr+a^{2}+Q^{2}\ ,
\label{eq:parametros_metrica_Kerr-Newman}
\end{equation}
and
\begin{equation}
\rho^{2}=r^{2}+a^{2}\cos^{2}\theta\ .
\end{equation}
We can write the metric tensor of the Kerr-Newman spacetime as
\begin{eqnarray*}
(g_{\sigma\tau})\!=\!
\end{eqnarray*}
\begin{eqnarray}
\left(\!
\begin{array}{cccc}
	\frac{\Delta\!-\!a^{2}\sin^{2}\theta}{\rho^{2}} & 0 & 0 & \frac{a\sin^{2}\theta\left[\!\left(\!r^{2}\!+\!a^{2}\!\right)\!-\!\Delta \!\right]}{\rho^{2}}\\
	0 & \!-\!\frac{\rho^{2}}{\Delta } & 0 & 0\\
	0 & 0 & \!-\!\rho^{2} & 0\\
	\frac{a\sin^{2}\theta\left[\!\left(\!r^{2}\!+\!a^{2}\!\right)\!-\!\Delta \!\right]}{\rho^{2}} & 0 & 0 & \!-\!\frac{\sin^{2}\theta\left[\!\left(\!r^{2}\!+\!a^{2}\!\right)^{2}\!-\!\Delta a^{2}\sin^{2}\theta\!\right]}{\rho^{2}}
\end{array}
\!\right)\ ,
\label{eq:metrica_matriz_Kerr-Newman}
\end{eqnarray}
from which we obtain
\begin{equation}
g \equiv \det(g_{\sigma\tau})=-\rho^{4}\sin^{2}\theta\ .
\label{eq:g_metrica_Kerr-Newman}
\end{equation}
Thus, the contravariant components of $g_{\sigma\tau}$ are given by
\begin{eqnarray}
(g^{\sigma\tau})\!=\!\left(\!
\begin{array}{cccc}
	\frac{\left(\!r^{2}\!+\!a^{2}\!\right)^{2}\!-\!\Delta a^{2}\sin^{2}\theta}{\rho^{2}\Delta  } & 0 & 0 & \frac{a\left[\! \left(\!r^{2}\!+\!a^{2}\!\right)\!-\!\Delta \!\right]}{\rho^{2}\Delta  }\\
	0 & \!-\!\frac{\Delta }{\rho^{2}} & 0 & 0\\
	0 & 0 & \!-\!\frac{1}{\rho^{2}} & 0\\
	\frac{a\left[\! \left(\!r^{2}\!+\!a^{2}\!\right)\!-\!\Delta \!\right]}{\rho^{2}\Delta  } & 0 & 0 & \!-\!\frac{\Delta \!-\! a^{2}\sin^{2}\theta}{\rho^{2}\Delta  \sin^{2}\theta}
\end{array}
\!\right)\ .
\label{eq:var_contra_metrica_Kerr-Newman}
\end{eqnarray}
From Eq.~(\ref{eq:parametros_metrica_Kerr-Newman}), we have that the horizon surface equation of the Kerr-Newman spacetime is obtained from the condition
\begin{equation}
\Delta=(r-r_{+})(r-r_{-})=0\ .
\label{eq:superficie_hor_Kerr-Newman}
\end{equation}
The solutions of Eq.~(\ref{eq:superficie_hor_Kerr-Newman}) are
\begin{equation}
r_{+}=M+\left[M^{2}-\left(a^{2}+Q^{2}\right)\right]^{1/2}\ ,
\label{eq:sol_padrao_Kerr-Newman_1}
\end{equation}
\begin{equation}
r_{-}=M-\left[M^{2}-\left(a^{2}+Q^{2}\right)\right]^{1/2}\ ,
\label{eq:sol_padrao_Kerr-Newman_2}
\end{equation}
and correspond to the event and Cauchy horizons of the Kerr-Newman black hole.

The gravitational acceleration on the black hole horizon surface $r_{+}$, and the Hawking radiation temperature are \cite{ChinJPhys.47.618}, respectively,
\begin{equation}
\kappa_{+} \equiv \frac{\Delta'(r_{+})}{2\left(r_{+}^{2}+a^{2}\right)}=\frac{r_{+}-r_{-}}{2\left(r_{+}^{2}+a^{2}\right)}\ ,
\label{eq:acel_grav_ext_Kerr-Newman}
\end{equation}
and
\begin{equation}
T_{+}=\frac{\kappa_{+}}{2\pi}\ .
\label{eq:temp_Hawking_Kerr-Newman}
\end{equation}
The thermodynamic quantities associated with the black hole, such as the entropy at event horizon, $S_{+}$, the dragging angular velocity of the exterior horizon, $\Omega_{+}$, the angular momentum, $J$, and the electric potential, $\Phi_{+}$, are given by
\begin{equation}
S_{+}=\pi\left(r_{+}^{2}+a^{2}\right)\ ,
\label{eq:entropia_Kerr-Newman}
\end{equation}
\begin{equation}
\Omega_{+}=\frac{a}{r_{+}^{2}+a^{2}}\ ,
\label{eq:vel_ang_Kerr-Newman}
\end{equation}
\begin{equation}
J=M a\ ,
\label{eq:parametros_Kerr-Newman_1}
\end{equation}
\begin{equation}
\Phi_{+}=\frac{Qr_{+}}{r_{+}^{2}+a^{2}}\ .
\label{eq:parametros_Kerr-Newman_2}
\end{equation}
These quantities obtained given by Eqs.~(\ref{eq:entropia_Kerr-Newman})-(\ref{eq:parametros_Kerr-Newman_2}) for the black hole event horizon satisfy the first law of thermodynamics 
\begin{equation}
dE=T_{+}\ dS_{+}+\Omega_{+}\ dJ+\Phi_{+}\ dQ\ .
\label{eq:1_lei_termo_Kerr-Newman}
\end{equation}
%
%
\section{The Klein-Gordon equation in a Kerr-Newman spacetime}
Now, let us consider the covariant Klein-Gordon equation for a charged massive scalar field in a curved spacetime and in the presence of an electromagnetic field. In this case, we can write the Klein-Gordon equation as
\begin{eqnarray}
& & \left[\frac{1}{\sqrt{-g}}\partial_{\sigma}\left(g^{\sigma\tau}\sqrt{-g}\partial_{\tau}\right)-ie(\partial_{\sigma}A^{\sigma})-2ieA^{\sigma}\partial_{\sigma}\right.\nonumber\\
& - & \left.\frac{ie}{\sqrt{-g}}A^{\sigma}(\partial_{\sigma}\sqrt{-g})-e^{2}A^{\sigma}A_{\sigma}+\mu_{0}^{2}\right]\Psi=0\ ,
\label{eq:Klein-Gordon_gauge_Kerr-Newman}
\end{eqnarray}
where $\mu_{0}$ is the mass of the scalar particle, and $e$ is the charge of the particle. The units $G \equiv c \equiv \hbar \equiv 1$ were chosen. The 4-vector electromagnetic potential is given by \cite{ChinAstronAstrophys.5.365}
\begin{equation}
A_{\sigma}=\frac{Qr}{\rho^{2}}\left(1,0,0,-a\sin^{2}\theta\right)\ .
\label{eq:potencial_EM_Kerr-Newman}
\end{equation}

Substituting Eqs.~(\ref{eq:g_metrica_Kerr-Newman}), (\ref{eq:var_contra_metrica_Kerr-Newman}), and (\ref{eq:potencial_EM_Kerr-Newman}) into Eq.~(\ref{eq:Klein-Gordon_gauge_Kerr-Newman}), we obtain
\begin{eqnarray}
& & \left\{\frac{1}{\Delta}\left[\left(r^{2}+a^{2}\right)^{2}-\Delta a^{2}\sin^{2}\theta\right]\frac{\partial^{2}}{\partial t^{2}}-\frac{\partial}{\partial r}\left(\Delta\frac{\partial}{\partial r}\right)\right.\nonumber\\
& - & \frac{1}{\sin\theta}\frac{\partial}{\partial\theta}\left(\sin\theta\frac{\partial}{\partial\theta}\right)-\frac{1}{\Delta \sin ^{2}\theta}\left(\Delta -a^{2}\sin^{2}\theta\right)\frac{\partial^{2}}{\partial\phi^{2}}\nonumber\\
& + & \frac{2  a}{\Delta }\left[\left(r^{2}+a^{2}\right)-\Delta \right]\frac{\partial^{2}}{\partial t\ \partial\phi}\nonumber\\
& + & 2ie\frac{Qr}{\Delta}\left[\left(r^{2}+a^{2}\right)\frac{\partial}{\partial t}+a\frac{\partial}{\partial\phi}\right]\nonumber\\
& + & \left.\mu_{0}^{2}\rho^{2}-e^{2}Q^{2}r^{2}\frac{1}{\Delta }\right\}\Psi=0\ .
\label{eq:mov_Kerr-Newman}
\end{eqnarray}
In order to solve Eq.~(\ref{eq:mov_Kerr-Newman}), we assume that its solution can be separated as follows
\begin{equation}
\Psi=\Psi(\mathbf{r},t)=R(r)S(\theta)\mbox{e}^{im\phi}\mbox{e}^{-i\omega t}\ .
\label{eq:separacao_variaveis}
\end{equation}
Substituting Eq.~(\ref{eq:separacao_variaveis}) into (\ref{eq:mov_Kerr-Newman}), we find that
\begin{eqnarray}
& & \frac{1}{\Delta}\left[\left(r^{2}+a^{2}\right)^{2}-\Delta a^{2}\sin^{2}\theta\right]\left(-\omega^{2}\right)-\frac{1}{R}\frac{d}{dr}\left(\Delta\frac{dR}{dr}\right)\nonumber\\
& - & \frac{1}{S}\frac{1}{\sin\theta}\frac{d}{d\theta}\left(\sin\theta\frac{dS}{d\theta}\right)\nonumber\\
& - & \frac{1}{\Delta \sin ^{2}\theta}\left(\Delta -a^{2}\sin^{2}\theta\right)\left(-m^{2}\right)\nonumber\\
& + & \frac{2  a}{\Delta }\left[\left(r^{2}+a^{2}\right)-\Delta \right](-i\omega)(im)\nonumber\\
& + & 2ie\frac{Qr}{\Delta}\left[\left(r^{2}+a^{2}\right)(-i\omega)+a(im)\right]\nonumber\\
& + & \mu_{0}^{2}\rho^{2}-e^{2}Q^{2}r^{2}\frac{1}{\Delta }=0\ .
\label{eq:mov_separavel}
\end{eqnarray}
This equation can be separated according to
\begin{eqnarray}
& & \frac{1}{\sin\theta}\frac{d}{d\theta}\left(\sin\theta\frac{dS}{d\theta}\right)\nonumber\\
& + & \left(\lambda_{lm}+c_{0}^{2}\cos^{2}\theta-\frac{m^{2}}{\sin^{2}\theta}\right)S=0\ ,
\label{eq:mov_angular}
\end{eqnarray}
where $c_{0}^{2}=a^{2}(\omega^{2}-\mu_{0}^{2})$, and
\begin{eqnarray}
& & \Delta\frac{d}{dr}\left(\Delta\frac{dR}{dr}\right)+\left\{\omega^{2}\right.\left(r^{2}+a^{2}\right)^{2}-4Ma\omega mr+2Q^{2}a\omega m\nonumber\\
& - & \mu_{0}^{2}r^{2}\Delta+m^{2}a^{2}-\left(\omega^{2}a^{2}+\lambda_{lm}\right)\Delta\nonumber\\
& - & \left.2eQr\left[\left(r^{2}+a^{2}\right)\omega-am\right]+e^{2}Q^{2}r^{2}\right\}R=0\ .
\label{eq:mov_radial_1}
\end{eqnarray}
Note that in the $e=0$ case, Eqs.~(\ref{eq:mov_angular}) and (\ref{eq:mov_radial_1}) reduce to Eqs.~(7) and (8), respectively, of our paper \cite{ClassQuantumGrav.31.045003}. In this separation, we have used the identity $\sin^{2}\theta=1-\cos^{2}\theta$ in the second term in square brackets of Eq.~(\ref{eq:mov_separavel}). However, as can be seen in \cite{ClassQuantumGrav.31.045003}, this separation gives us solutions of the radial equation in terms of the confluent Heun functions, in which its parameters contain terms proportional to $M$, the mass of the black hole, which is a result not suitable to study the Hawking radiation.

Then, for convenience, instead of adopting the procedure of our paper \cite{ClassQuantumGrav.31.045003}, we will keep $\sin^{2}\theta$ in the second term in square brackets of Eq.~(\ref{eq:mov_separavel}). Thus, we can separate this equation according to
\begin{eqnarray}
& & \frac{1}{\sin\theta}\frac{d}{d\theta}\left(\sin\theta\frac{dS}{d\theta}\right)\nonumber\\
& + & \left[-\left(\omega a\sin\theta-\frac{m}{\sin\theta}\right)^{2}-\mu_{0}^{2}a^{2}\cos^{2}\theta+\lambda\right]S=0\ ,
\label{eq:mov_angular_Kerr-Newman}
\end{eqnarray}
\begin{eqnarray}
& & \frac{d}{dr}\left(\Delta\frac{dR}{dr}\right)+\{-\left(\lambda+\mu_{0}^{2}r^{2}\right)\nonumber\\
& + & \frac{1}{\Delta}\left[\omega\left(r^{2}+a^{2}\right)-am-eQr\right]^{2}\}R=0\ ,
\label{eq:mov_radial_1_Kerr-Newman}
\end{eqnarray}
where $\lambda$ is the separation constant, $\omega$ is the energy of the particle, and $m$ is the azimuthal quantum number. In what follows we will solve the angular and radial parts of the Klein-Gordon equation, whose solutions are given in terms of the confluent Heun functions and use these solutions to study the Hawking radiation.
%
%
\subsection{Angular equation}
Now, let us obtain the exact and general solution for the angular part of the Klein-Gordon equation given by Eq.~(\ref{eq:mov_angular_Kerr-Newman}), which can be rewritten as
\begin{eqnarray}
& & \frac{1}{\sin\theta}\frac{d}{d\theta}\left(\sin\theta\frac{dS}{d\theta}\right)\nonumber\\
& + & \left(\Lambda_{lm}+c_{0}^{2}\cos^{2}\theta-\frac{m^{2}}{\sin^{2}\theta}\right)S=0\ ,
\label{eq:mov_angular_Kerr-Newman_2}
\end{eqnarray}
where $c_{0}^{2}=a^{2}(\omega^{2}-\mu_{0}^{2})$, and the constant $\Lambda_{lm}$ is defined by
\begin{equation}
\Lambda_{lm} \equiv \lambda-a^{2}\omega^{2}+2a \omega m\ .
\label{eq:cte_sep_Kerr-Newman_2}
\end{equation}

In the literature, the solutions of Eq.~(\ref{eq:mov_angular_Kerr-Newman_2}) are the oblate spheroidal harmonic functions $S_{lm}(ic_{0},\cos\theta)$ with eigenvalues $\Lambda_{lm}$, where $l,m$ are integers such that $|m|\leq l$  \cite{PhysRevD.12.2963,Morse:1953}. We will show that the solutions of Eq.~(\ref{eq:mov_angular_Kerr-Newman_2}) can be expressed in terms of the confluent Heun functions.

To do this, let us rewrite Eq.~(\ref{eq:mov_angular_Kerr-Newman_2}) in the form which resembles a Heun equation by defining a new angular coordinate, $x$, such that
\begin{equation}
x=\cos^{2}\theta\ .
\label{eq:coord_ang_Kerr-Newman_2}
\end{equation}
Thus, we can write Eq.~(\ref{eq:mov_angular_Kerr-Newman_2}) as
\begin{eqnarray}
& & \frac{d^{2}S}{dx^{2}}+\left(\frac{1/2}{x}+\frac{1}{x-1}\right)\frac{dS}{dx}\nonumber\\
& + & \frac{1}{x(x-1)}\left[-\frac{c_{0}^{2}x}{4}-\frac{\Lambda_{lm}}{4}-\frac{m^{2}}{4(x-1)}\right]S=0\ ,
\label{eq:mov_angular_x_Kerr-Newman_2}
\end{eqnarray}
which has singularities at $x=0,1$ and $x=\infty$. Equation (\ref{eq:mov_angular_x_Kerr-Newman_2}) can also be written as
\begin{eqnarray}
& & \frac{d^{2}S}{dx^{2}}+\left(\frac{1/2}{x}+\frac{1}{x-1}\right)\frac{dS}{dx}\nonumber\\
& + & \left[\frac{A_{1}}{x}+\frac{A_{2}}{x-1}+\frac{A_{3}}{(x-1)^{2}}\right]S=0\ ,
\label{eq:mov_angular_x_Kerr-Newman_3}
\end{eqnarray}
where the coefficients $A_{1}$, $A_{2}$, and $A_{3}$ are given by:
\begin{equation}
A_{1}=\frac{\Lambda_{lm}-m^{2}}{4}\ ;
\label{eq:A1_mov_angular_x_Kerr-Newman_3}
\end{equation}
\begin{equation}
A_{2}=-\frac{\left(\Lambda_{lm}-m^{2}+c_{0}^{2}\right)}{4}\ ;
\label{eq:A2_mov_angular_x_Kerr-Newman_3}
\end{equation}
\begin{equation}
A_{3}=-\frac{m^{2}}{4}\ .
\label{eq:A3_mov_angular_x_Kerr-Newman_3}
\end{equation}

Defining a new function, $S(x)$, by $S(x)=Z(x)x^{-1/4}(x-1)^{-1/2}$, we can write Eq.~(\ref{eq:mov_angular_x_Kerr-Newman_3}) in the following form
\begin{equation}
\frac{d^{2}Z}{dx^{2}}+\left[\frac{B_{2}}{x}+\frac{B_{3}}{x-1}+\frac{B_{4}}{x^{2}}+\frac{B_{5}}{(x-1)^{2}}\right]Z=0\ ,
\label{eq:mov_angular_x_heun}
\end{equation}
where the coefficients $B_{2}$, $B_{3}$, $B_{4}$, and $B_{5}$ are given by:
\begin{equation}
B_{2}=\frac{1+4A_{1}}{4}\ ;
\label{eq:B2_mov_angular_x_Kerr-Newman_3}
\end{equation}
\begin{equation}
B_{3}=\frac{-1+4A_{2}}{4}\ ;
\label{eq:B3_mov_angular_x_Kerr-Newman_3}
\end{equation}
\begin{equation}
B_{4}=\frac{3}{16}\ ;
\label{eq:B4_mov_angular_x_Kerr-Newman_3}
\end{equation}
\begin{equation}
B_{5}=\frac{1+4A_{3}}{4}\ .
\label{eq:B5_mov_angular_x_Kerr-Newman_3}
\end{equation}

Now, consider an equation in the standard form
\begin{equation}
\frac{d^{2}U}{dz^{2}}+p(z)\frac{dU}{dz}+q(z)U=0\ .
\label{eq:EDO_forma_padrao}
\end{equation}
Changing the function $U(z)$, using the relation
\begin{equation}
U(z)=Z(z)\mbox{e}^{-\frac{1}{2}\int p(z)dz}\ ,
\label{eq:U}
\end{equation}
Eq.~(\ref{eq:EDO_forma_padrao}) turns into the normal form
\begin{equation}
\frac{d^2Z}{dz^2}+I(z)Z=0\ ,
\label{eq:EDO_forma_normal}
\end{equation}
where
\begin{equation}
I(z)=q(z)-\frac{1}{2}\frac{dp(z)}{dz}-\frac{1}{4}\left[p(z)\right]^2\ .
\label{eq:I}
\end{equation}
Now, let us consider the confluent Heun equation \cite{JPhysAMathTheor.43.035203}
\begin{eqnarray}
& & \frac{d^{2}U}{dz^{2}}+\left(\alpha+\frac{\beta+1}{z}+\frac{\gamma+1}{z-1}\right)\frac{dU}{dz}\nonumber\\
& + & \left(\frac{\mu}{z}+\frac{\nu}{z-1}\right)U=0\ ,
\label{eq:Heun_confluente_forma_canonica}
\end{eqnarray}
where $U(z)=\mbox{HeunC}(\alpha,\beta,\gamma,\delta,\eta;z)$ are the confluent Heun functions, with the parameters $\alpha$, $\beta$, $\gamma$, $\delta$ and $\eta$, which are related to $\mu$ and $\nu$ by the following expressions
\begin{equation}
\mu=\frac{1}{2}(\alpha-\beta-\gamma+\alpha\beta-\beta\gamma)-\eta\ ,
\label{eq:mu_Heun_conlfuente_2}
\end{equation}
\begin{equation}
\nu=\frac{1}{2}(\alpha+\beta+\gamma+\alpha\gamma+\beta\gamma)+\delta+\eta\ ,
\label{eq:nu_Heun_conlfuente_2}
\end{equation}
according to the standard package of the \textbf{Maple}\texttrademark \textbf{17}. Using Eqs.~(\ref{eq:EDO_forma_padrao})-(\ref{eq:I}), we can write Eq.~(\ref{eq:Heun_confluente_forma_canonica}) in the normal form as \cite{Ronveaux:1995,Slavyanov:2000}
\begin{equation}
\frac{d^{2}Z}{dz^{2}}+\left[D_{1}+\frac{D_{2}}{z}+\frac{D_{3}}{z-1}+\frac{D_{4}}{z^{2}}+\frac{D_{5}}{(z-1)^{2}}\right]Z=0\ ,
\label{eq:Heun_confluente_forma_normal}
\end{equation}
where the coefficients $D_{1}$, $D_{2}$, $D_{3}$, $D_{4}$, and $D_{5}$ are given by:
\begin{equation}
D_{1} \equiv -\frac{1}{4}\alpha^{2}\ ;
\label{eq:D1_Heun_confluente_forma_normal}
\end{equation}
\begin{equation}
D_{2} \equiv\frac{1}{2}(1-2\eta)\ ;
\label{eq:D2_Heun_confluente_forma_normal}
\end{equation}
\begin{equation}
D_{3} \equiv \frac{1}{2}(-1+2\delta+2\eta)\ ;
\label{eq:D3_Heun_confluente_forma_normal}
\end{equation}
\begin{equation}
D_{4} \equiv \frac{1}{4}(1-\beta^{2})\ ;
\label{eq:D4_Heun_confluente_forma_normal}
\end{equation}
\begin{equation}
D_{5} \equiv \frac{1}{4}(1-\gamma^{2})\ .
\label{eq:D5_Heun_confluente_forma_normal}
\end{equation}
The angular part of the Klein-Gordon equation for a charged massive scalar particle in the Kerr-Newman spacetime in the exterior region to the event horizon, given by (\ref{eq:mov_angular_Kerr-Newman}), can be written as (\ref{eq:Heun_confluente_forma_normal}), and therefore, its solution is given by
\begin{equation}
Z(z)=U(z)\mbox{e}^{\frac{1}{2}\int\left(\alpha+\frac{\beta+1}{z}+\frac{\gamma+1}{z-1}\right)dz}\ ,
\label{eq:solZ}
\end{equation}
where $U(z)$ is a solution of the confluent Heun equation (\ref{eq:Heun_confluente_forma_canonica}), and the parameters $\alpha$, $\beta$, $\gamma$, $\delta$, and $\eta$ are obtained from the following relations:
\begin{equation}
-\frac{1}{4}\alpha^{2}=0\ ;
\label{eq:D1_mov_angular_x_heun}
\end{equation}
\begin{equation}
\frac{1}{2}(1-2\eta)=\frac{1+4A_{1}}{4}\ ;
\label{eq:D2_mov_angular_x_heun}
\end{equation}
\begin{equation}
\frac{1}{2}(-1+2\delta+2\eta)=\frac{-1+4A_{2}}{4}\ ;
\label{eq:D3_mov_angular_x_heun}
\end{equation}
\begin{equation}
\frac{1}{4}(1-\beta^{2})=\frac{3}{16}\ ;
\label{eq:D4_mov_angular_x_heun}
\end{equation}
\begin{equation}
\frac{1}{4}(1-\gamma^{2})=\frac{1+4A_{3}}{4}\ .
\label{eq:D5_mov_angular_x_heun}
\end{equation}
Thus, from the above relations, we find that:
\begin{equation}
\alpha=0\ ;
\label{eq:alpha_mov_angular_x_HeunC_Kerr-Newman_3}
\end{equation}
\begin{equation}
\beta=\frac{1}{2}\ ;
\label{eq:beta_mov_angular_x_HeunC_Kerr-Newman_3}
\end{equation}
\begin{equation}
\gamma=m\ ;
\label{eq:gamma_mov_angular_x_HeunC_Kerr-Newman_3}
\end{equation}
\begin{eqnarray}
\delta & = & -\frac{c_{0}^{2}}{4}\nonumber\\
& = & \frac{1}{4}\left(a^{2}\mu_{0}^{2}-a^{2}\omega^{2}\right)\ ;
\label{eq:delta_mov_angular_x_HeunC_Kerr-Newman_3}
\end{eqnarray}
\begin{eqnarray}
\eta & = & \frac{1}{4}\left(1+m^2-\Lambda_{lm}\right)\nonumber\\
& = & \frac{1}{4}\left(1+m^2-\lambda+a^{2}\omega^{2}-2a \omega m\right)\ .
\label{eq:eta_mov_angular_x_HeunC_Kerr-Newman_3}
\end{eqnarray}
The general solution of Eq.~(\ref{eq:mov_angular_x_Kerr-Newman_3}) over the entire range $0 \leq x < \infty$ is obtained with the use of Eq.~(\ref{eq:solZ}). It is given by
\begin{eqnarray}
S(x) & = & (x-1)^{\frac{1}{2}\gamma}x^{\frac{1}{2}\left(\frac{1}{2}+\beta\right)}\nonumber\\
& \times & \{C_{1}\ \mbox{HeunC}(\alpha,\beta,\gamma,\delta,\eta;x)\nonumber\\
& + & C_{2}\ x^{-\beta}\ \mbox{HeunC}(\alpha,-\beta,\gamma,\delta,\eta;x)\}\ ,
\label{eq:solucao_geral_mov_angular_x_Kerr-Newman_3}
\end{eqnarray}
where $C_{1}$ and $C_{2}$ are constants, and the parameters $\alpha$, $\beta$, $\gamma$, $\delta$, and $\eta$ are fixed by relations (\ref{eq:alpha_mov_angular_x_HeunC_Kerr-Newman_3})-(\ref{eq:eta_mov_angular_x_HeunC_Kerr-Newman_3}). These two functions form linearly independent solutions of the confluent Heun differential equation provided $\beta$ is not integer.
%
%
\subsection{Radial equation}
Now, let us obtain the exact and general solution for the radial part of the Klein-Gordon equation given by Eq.~(\ref{eq:mov_radial_1_Kerr-Newman}). Indeed, we can generalize the results obtained in our previous work \cite{ClassQuantumGrav.31.045003}.

Therefore, to solve the radial part of the Klein-Gordon equation, we use Eq.~(\ref{eq:superficie_hor_Kerr-Newman}) and write down Eq.~(\ref{eq:mov_radial_1_Kerr-Newman}) as
\begin{eqnarray}
& & \frac{d^{2}R}{dr^{2}}+\left(\frac{1}{r-r_{+}}+\frac{1}{r-r_{-}}\right)\frac{dR}{dr}\nonumber\\
& + & \frac{1}{(r-r_{+})(r-r_{-})}\left\{\right.r^{2}\left(\omega^{2}-\mu^{2}\right)\nonumber\\
& + & r\left[\omega^{2}(r_{+}+r_{-})-2eQ\omega\right]\nonumber\\
& + & e^{2}Q^{2}-\lambda-2am\omega-2eQ\omega(r_{+}+r_{-})+2a^{2}\omega^{2}\nonumber\\
& + & \omega^{2}\left(r_{+}^{2}+r_{+}r_{-}+r_{-}^{2}\right)\nonumber\\
& + & \frac{\left(-a m-e Q r_{+}+a^2 \omega +r_{+}^2 \omega \right)^2}{(r-r_{+}) (r_{+}-r_{-})}\nonumber\\
& - & \frac{\left(-a m-e Q r_{-}+a^2 \omega +r_{-}^2 \omega \right)^2}{(r-r_{-}) (r_{+}-r_{-})}\left.\right\}R=0\ .
\label{eq:mov_radial_2_Kerr-Newman}
\end{eqnarray}
This equation has singularities at $r=(a_{1},a_{2})=(r_{+},r_{-})$, and at $r=\infty$. The transformation of (\ref{eq:mov_radial_2_Kerr-Newman}) to a Heun-type equation is achieved by setting
\begin{equation}
x=\frac{r-a_{1}}{a_{2}-a_{1}}=\frac{r-r_{+}}{r_{-}-r_{+}}\ .
\label{eq:homog_subs_radial}
\end{equation}
Thus, we can written Eq.~(\ref{eq:mov_radial_2_Kerr-Newman}) as
\begin{eqnarray}
& & \frac{d^{2}R}{dx^{2}}+\left(\frac{1}{x}+\frac{1}{x-1}\right)\frac{dR}{dx}\nonumber\\
& + & \left[D_{1}+\frac{D_{2}}{x}+\frac{D_{3}}{x-1}+\frac{D_{4}}{x^{2}}+\frac{D_{5}}{(x-1)^{2}}\right]R=0\ ,
\label{eq:mov_radial_2_Kerr-Newman_x}
\end{eqnarray}
where the coefficients $D_{1}$, $D_{2}$, $D_{3}$, $D_{4}$, and $D_{5}$ are given by:
\begin{eqnarray}
D_{1} & = & \left(\omega^{2}-\mu^{2}\right)(r_{+}-r_{-})^{2}\ ;
\label{eq:D1_mov_radial_x_normal}
\end{eqnarray}
\begin{eqnarray}
D_{2} & = & \frac{2 a^4 \omega ^2-4 a^3 m \omega -2 a^2 e Q r_{+} \omega -2 a^2 e Q r_{-} \omega +2 a^2 m^2}{(r_{+}-r_{-})^2}\nonumber\\
& + & \frac{4 a^2 r_{+} r_{-} \omega ^2+2 a e m Q r_{+}+2 a e m Q r_{-}-4 a m r_{+} r_{-} \omega }{(r_{+}-r_{-})^2}\nonumber\\
& + & \frac{2 e^2 Q^2 r_{+} r_{-}+2 e Q r_{+}^3 \omega -6 e Q r_{+}^2 r_{-} \omega +\mu ^2 r_{+}^4}{(r_{+}-r_{-})^2}\nonumber\\
& + & \frac{-2 r_{+}^4 \omega ^2-2 \mu ^2 r_{+}^3 r_{-}+4 r_{+}^3 r_{-} \omega ^2+\lambda  r_{+}^2+\mu ^2 r_{+}^2 r_{-}^2}{(r_{+}-r_{-})^2}\nonumber\\
& + & \frac{-2 \lambda  r_{+} r_{-}+\lambda  r_{-}^2}{(r_{+}-r_{-})^2}\ ;
\label{eq:D2_mov_radial_x_normal}
\end{eqnarray}
\begin{eqnarray}
D_{3} & = & \frac{-2 a^4 \omega ^2+4 a^3 m \omega +2 a^2 e Q r_{+} \omega +2 a^2 e Q r_{-} \omega }{(r_{+}-r_{-})^2}\nonumber\\
& + & \frac{-2 a^2 m^2-4 a^2 r_{+} r_{-} \omega ^2-2 a e m Q r_{+}-2 a e m Q r_{+}}{(r_{+}-r_{-})^2}\nonumber\\
& + & \frac{-2 a e m Q r_{-}+4 a m r_{+} r_{-} \omega -2 e^2 Q^2 r_{+} r_{-}}{(r_{+}-r_{-})^2}\nonumber\\
& + & \frac{6 e Q r_{+} r_{-}^2 \omega -2 e Q r_{-}^3 \omega -\lambda  r_{+}^2-\mu ^2 r_{+}^2 r_{-}^2+2 \mu ^2 r_{+} r_{-}^3}{(r_{+}-r_{-})^2}\nonumber\\
& + & \frac{-4 r_{+} r_{-}^3 \omega ^2+2 \lambda  r_{+} r_{-}-\mu ^2 r_{-}^4+2 r_{-}^4 \omega ^2-\lambda  r_{-}^2}{(r_{+}-r_{-})^2}\ ;
\label{eq:D3_mov_radial_x_normal}
\end{eqnarray}
\begin{eqnarray}
D_{4} & = & \frac{a^4 \omega ^2-2 a^3 m \omega -2 a^2 e Q r_{+} \omega +a^2 m^2+2 a^2 r_{+}^2 \omega ^2}{(r_{+}-r_{-})^2}\nonumber\\
& + & \frac{2 a e m Q r_{+}-2 a m r_{+}^2 \omega +e^2 Q^2 r_{+}^2+e^2 Q^2 r_{+}^2}{(r_{+}-r_{-})^2}\nonumber\\
& + & \frac{-2 e Q r_{+}^3 \omega +r_{+}^4 \omega ^2}{(r_{+}-r_{-})^2}\ ;
\label{eq:D4_mov_radial_x_normal}
\end{eqnarray}
\begin{eqnarray}
D_{5} & = & \frac{a^4 \omega ^2-2 a^3 m \omega -2 a^2 e Q r_{-} \omega +a^2 m^2+2 a^2 r_{-}^2 \omega ^2}{(r_{+}-r_{-})^2}\nonumber\\
& + & \frac{2 a e m Q r_{-}-2 a m r_{-}^2 \omega +e^2 Q^2 r_{-}^2+e^2 Q^2 r_{-}^2}{(r_{+}-r_{-})^2}\nonumber\\
& + & \frac{-2 e Q r_{-}^3 \omega +r_{-}^4 \omega ^2}{(r_{+}-r_{-})^2}\ .
\label{eq:D5_mov_radial_x_normal}
\end{eqnarray}

Defining a new function, $R(x)$, by $R(x)=Z(x)[x(x-1)]^{-1/2}$, we can write Eq.~(\ref{eq:mov_radial_2_Kerr-Newman_x}) in the following form
\begin{equation}
\frac{d^{2}Z}{dx^{2}}+\left[E_{1}+\frac{E_{2}}{x}+\frac{E_{3}}{x-1}+\frac{E_{4}}{x^{2}}+\frac{E_{5}}{(x-1)^{2}}\right]Z=0\ ,
\label{eq:mov_radial_x_heun}
\end{equation}
with the coefficients $E_{1}$, $E_{2}$, $E_{3}$, $E_{4}$, and $E_{5}$ given by:
\begin{eqnarray}
E_{1} & = & \left(\omega^{2}-\mu^{2}\right)(r_{+}-r_{-})^{2}\ ;
\label{eq:E1_mov_radial_x_normal}
\end{eqnarray}
\begin{eqnarray}
E_{2} & = & \frac{\left(4 e^{2} Q^{2}-8 a m \omega+8 a^{2} \omega^{2}\right) r_{+} r_{-} }{2 \left(r_{+}-r_{-}\right)^{2}}\nonumber\\
& + & \frac{-4 e Q r_{-} \omega\left(a^{2}+3r_{+}^{2}\right)-4 \omega ^{2}r_{+}^{3}\left(r_{+} -2 r_{-}\right)}{2 \left(r_{+}-r_{-}\right)^{2}}\nonumber\\
& + & \frac{4 a^{2}\left(m-a \omega\right)^{2}+4 a e m Q \left(r_{+}+r_{-}\right)}{2 \left(r_{+}-r_{-}\right)^{2}}\nonumber\\
& + & \frac{\left(2 \lambda+1+2 r_{+}^{2} \mu ^{2}\right)\left(r_{+}-r_{-}\right)^{2}}{2 \left(r_{+}-r_{-}\right)^{2}}\nonumber\\
& + & \frac{4 e Q r_{+} \omega\left(r_{+}^{2}-a^{2}\right)}{2 \left(r_{+}-r_{-}\right)^{2}}\ ;
\label{eq:E2_mov_radial_x_normal}
\end{eqnarray}
\begin{eqnarray}
E_{3} & = & \frac{-\left(4 e^{2} Q^{2}-8 a m \omega+8 a^{2} \omega ^{2}\right)r_{+} r_{-} }{2 \left(r_{+}-r_{-}\right)^{2}}\nonumber\\
& + & \frac{4 e Q r_{+} \omega\left(a^{2}+3r_{-}^{2}\right)+4 \omega ^{2}r_{-}^{3}\left(r_{-} -2 r_{+}\right)}{2 \left(r_{+}-r_{-}\right)^{2}}\nonumber\\
& - & \frac{4 a^{2}\left(m-a \omega\right)^{2}+4 a e m Q \left(r_{+}+r_{-}\right) }{2 \left(r_{+}-r_{-}\right)^{2}}\nonumber\\
& - & \frac{\left(2 \lambda+1+2 r_{-}^{2} \mu ^{2}\right)\left(r_{+}-r_{-}\right)^{2}}{2 \left(r_{+}-r_{-}\right)^{2}}\nonumber\\
& - & \frac{4 e Q r_{-} \omega\left(r_{-}^{2}-a^{2}\right)}{2 \left(r_{+}-r_{-}\right)^{2}}\ ;
\label{eq:E3_mov_radial_x_normal}
\end{eqnarray}
\begin{eqnarray}
E_{4} & = & \frac{\left(4 e^{2} Q^{2}-8 a m \omega+8 a^{2} \omega ^{2}\right) r_{+}^{2} -8 e Q r_{+} \omega\left(a^{2}+r_{+}^{2}\right) }{4 \left(r_{+}-r_{-}\right)^{2}}\nonumber\\
& + & \frac{4 r_{+}^{4} \omega ^{2}+4 a^{2}\left(m-a \omega\right)^{2}+8 a e m Q r_{+}}{4 \left(r_{+}-r_{-}\right)^{2}}\nonumber\\
& + & \frac{1}{4}\ ;
\label{eq:E4_mov_radial_x_normal}
\end{eqnarray}
\begin{eqnarray}
E_{5} & = & \frac{\left(4 e^{2} Q^{2}-8 a m \omega+8 a^{2} \omega ^{2}\right) r_{-}^{2} -8 e Q r_{-} \omega\left(a^{2}+r_{-}^{2}\right) }{4 \left(r_{+}-r_{-}\right)^{2}}\nonumber\\
& + & \frac{4 r_{-}^{4} \omega ^{2}+4 a^{2}\left(m-a \omega\right)^{2}+8 a e m Q r_{-}}{4 \left(r_{+}-r_{-}\right)^{2}}\nonumber\\
& + & \frac{1}{4}\ .
\label{eq:E5_mov_radial_x_normal}
\end{eqnarray}
The radial part of the Klein-Gordon equation for a charged massive scalar particle in the Kerr-Newman spacetime in the exterior region of event horizon, given by Eq.~(\ref{eq:mov_radial_1_Kerr-Newman}), can be written as (\ref{eq:Heun_confluente_forma_normal}). Therefore, the general solution of Eq.~(\ref{eq:mov_radial_2_Kerr-Newman_x}) over the entire range $0 \leq x < \infty$ is obtained with the use of Eq.~(\ref{eq:solZ}). It is given by
\begin{eqnarray}
R(x) & = & \mbox{e}^{\frac{1}{2}\alpha x}(x-1)^{\frac{1}{2}\gamma}x^{\frac{1}{2}\beta}\nonumber\\
& \times & \{C_{1}\ \mbox{HeunC}(\alpha,\beta,\gamma,\delta,\eta;x)\nonumber\\
& + & C_{2}\ x^{-\beta}\ \mbox{HeunC}(\alpha,-\beta,\gamma,\delta,\eta;x)\}\ ,
\label{eq:solucao_geral_radial_Kerr-Newman_gauge}
\end{eqnarray}
where $C_{1}$ and $C_{2}$ are constants, and the parameters $\alpha$, $\beta$, $\gamma$, $\delta$, and $\eta$ are given by:
\begin{equation}
\alpha=2(r_{+}-r_{-})\left(\mu^{2}-\omega^{2}\right)^{1/2}\ ;
\label{eq:alpha_radial_HeunC_Kerr-Newman}
\end{equation}
\begin{equation}
\beta=\frac{2 i \left[\omega (r_{+}^{2}+a^{2}) -a m-e Q r_{+}\right]}{r_{+}-r_{-}}\ ;
\label{eq:beta_radial_HeunC_Kerr-Newman}
\end{equation}
\begin{equation}
\gamma=\frac{2 i \left[\omega (r_{-}^{2}+a^{2}) -a m-e Q r_{-}\right]}{r_{+}-r_{-}}\ ;
\label{eq:gamma_radial_HeunC_Kerr-Newman}
\end{equation}
\begin{equation}
\delta=(r_{+}-r_{-})\left[2 e Q \omega +(r_{+}+r_{-})\left(\mu ^{2}-2 \omega ^{2}\right)\right]\ ;
\label{eq:delta_radial_HeunC_Kerr-Newman}
\end{equation}
\begin{eqnarray}
\eta & = & \frac{-2 a^{2}\left(m-a \omega\right)^{2}-2 a e m Q (r_{+}+r_{-})}{(r_{+}-r_{-})^{2}}\nonumber\\
& + & \frac{- \left(\lambda +r_{+}^{2} \mu ^{2}\right)(r_{+}-r_{-})^{2}}{(r_{+}-r_{-})^{2}}\nonumber\\
& + & \frac{-(2 e^{2} Q^{2}-4 a m \omega+4 a^{2} \omega^{2}) r_{+} r_{-} }{(r_{+}-r_{-})^{2}}\nonumber\\
& + & \frac{-2 e Q r_{-} \omega\left(a^{2}-3r_{+}^{2}\right)}{(r_{+}-r_{-})^{2}}\nonumber\\
& + & \frac{2 \omega ^{2} r_{+}^3 (r_{+}-2 r_{-})-2 e Q r_{+} \omega\left(r_{+}^{2}-a^{2}\right)}{(r_{+}-r_{-})^{2}}\ .
\label{eq:eta_radial_HeunC_Kerr-Newman}
\end{eqnarray}
These two functions form linearly independent solutions of the confluent Heun dif\-fe\-ren\-tial equation provided $\beta$ is not integer. However, there is not any specific physical reason to impose that $\beta$ should be integer. If we consider the expansion in power series of the confluent Heun functions with respect to the independent variable $x$ in a vicinity of the regular singular point $x=0$ \cite{Ronveaux:1995}, we can write
\begin{eqnarray}
\mbox{HeunC}(\alpha,\beta,\gamma,\delta,\eta;x) & = & 1\nonumber\\
& + & \frac{1}{2}\frac{1}{(\beta+1)}(-\alpha\beta+\beta\gamma+2\eta-\alpha\nonumber\\
& + & \beta+\gamma)x\nonumber\\
& + & \frac{1}{8}\frac{1}{(\beta+1)(\beta+2)}\left(\alpha^{2}\beta^{2}\right.\nonumber\\
& - & 2\alpha\beta^{2}\gamma+\beta^{2}\gamma^{2}-4\eta\alpha\beta\nonumber\\
& + & 4\eta\beta\gamma+4\alpha^{2}\beta-2\alpha\beta^{2}-6\alpha\beta\gamma\nonumber\\
& + & 4\beta^{2}\gamma+4\beta\gamma^{2}+4\eta^{2}-8\eta\alpha\nonumber\\
& + & 8\eta\beta+8\eta\gamma+3\alpha^{2}-4\alpha\beta\nonumber\\
& - & 4\alpha\gamma+3\beta^{2}+4\beta\delta+10\beta\gamma\nonumber\\
& + & \left.3\gamma^{2}+8\eta+4\beta+4\delta+4\gamma\right)x^2\nonumber\\
& + & ...\ ,
\label{eq:serie_HeunC_todo_x}
\end{eqnarray}
which is a useful form to be used in the discussion of Hawking radiation.
%
%
\section{Hawking radiation}
We will consider the charged massive scalar field near the horizon in order to discuss the Hawking radiation.

From Eqs.~(\ref{eq:homog_subs_radial}) and (\ref{eq:serie_HeunC_todo_x}) we can see that the radial solution given by Eq.~(\ref{eq:solucao_geral_radial_Kerr-Newman_gauge}), near the exterior event horizon, that is, when $r \rightarrow r_{+} \Rightarrow x \rightarrow 0$, behave asymptotically as
\begin{equation}
R(r) \sim C_{1}\ (r-r_{+})^{\beta/2}+C_{2}\ (r-r_{+})^{-\beta/2}\ ,
\label{eq:exp_0_solucao_geral_radial_Kerr-Newman_gauge}
\end{equation}
where we are considering contributions only of the first term in the expansion, and all constants are included in $C_{1}$ and $C_{2}$. Thus, considering the time factor, near the black hole event horizon $r_{+}$, this solution is
\begin{equation}
\Psi=\mbox{e}^{-i \omega t}(r-r_{+})^{\pm\beta/2}\ .
\label{eq:sol_onda_radial_Kerr-Newman_gauge}
\end{equation}
From Eq.~(\ref{eq:beta_radial_HeunC_Kerr-Newman}), for the parameter $\beta$, we obtain
\begin{equation}
\frac{\beta}{2}=i\left[\omega\frac{r_{+}^{2}+a^{2}}{r_{+}-r_{-}}-\left(m\frac{a}{r_{+}-r_{-}}+e\frac{Q r_{+}}{r_{+}-r_{-}}\right)\right]\ .
\label{eq:beta/2_solucao_geral_radial_Kerr-Newman_gauge}
\end{equation}
From Eq.~(\ref{eq:acel_grav_ext_Kerr-Newman}), we have
\begin{equation}
\frac{1}{2\kappa_{+}}=\frac{r_{+}^{2}+a^{2}}{r_{+}-r_{-}}\ \Leftrightarrow\ r_{+}-r_{-}=2\kappa_{+}\left(r_{+}^{2}+a^{2}\right)\ .
\label{eq:rel_2_solucao_geral_radial_Kerr-Newman_gauge}
\end{equation}
Then, substituting Eq.~(\ref{eq:rel_2_solucao_geral_radial_Kerr-Newman_gauge}) into Eq.~(\ref{eq:beta/2_solucao_geral_radial_Kerr-Newman_gauge}), we get
\begin{eqnarray}
\frac{\beta}{2} & = & \frac{i}{2\kappa_{+}}\left[\omega-\left(m\frac{a}{r_{+}^{2}+a^{2}}+e\frac{Q r_{+}}{r_{+}^{2}+a^{2}}\right)\right]\nonumber\\
& = & \frac{i}{2\kappa_{+}}[\omega-(m\Omega_{+}+e\Phi_{+})]\nonumber\\
& = & \frac{i}{2\kappa_{+}}(\omega-\omega_{0})\ ,
\label{eq:expoente_rad_Hawking_Kerr-Newman_gauge}
\end{eqnarray}
where $\omega_{0}=m\Omega_{+}+e\Phi_{+}$.

Therefore, on the black hole exterior horizon surface the ingoing and outgoing wave solutions are
\begin{equation}
\Psi_{in}=\mbox{e}^{-i \omega t}(r-r_{+})^{-\frac{i}{2\kappa_{+}}(\omega-\omega_{0})}\ ,
\label{eq:sol_in_1_Kerr-Newman_gauge}
\end{equation}
\begin{equation}
\Psi_{out}(r>r_{+})=\mbox{e}^{-i \omega t}(r-r_{+})^{\frac{i}{2\kappa_{+}}(\omega-\omega_{0})}\ .
\label{eq:sol_out_2_Kerr-Newman_gauge}
\end{equation}
Thus, we obtained the solutions for ingoing and outgoing waves near to the exterior horizon $r_{+}$ of a Kerr-Newman black hole. These solutions for the scalar fields near the horizon will be useful to investigate Hawking radiation of charged massive scalar particles. It is worth calling attention to the fact that we are using the analytical solution of the radial part of the Klein-Gordon equation in the spacetime under consideration, differently from the calculations usually done in the literature \cite{PhysLettB.618.14,ChinJPhys.47.618}.

For consistency and completeness, we will show that our solution is exactly the same one obtained by other methods \cite{PhysLettB.618.14,ChinJPhys.47.618}, unless a change of variable. Using the definitions of the tortoise and Eddington-Finkelstein coordinates, given by \cite{PhysRevD.14.332}
\begin{equation}
dr_{*}=\frac{1}{\Delta}\left(r^{2}+a^{2}\right)dr\ \Rightarrow
\label{eq:coord_tortoise_1.1}
\end{equation}
\begin{equation}
\ln(r-r_{+})=\frac{1}{r_{+}^{2}+a^{2}}\left.\frac{d\Delta}{dr}\right|_{r=r_{+}}r_{*}=2\kappa_{+}r_{*}\ ,
\label{eq:coord_tortoise_1}
\end{equation}
\begin{equation}
\hat{r}=\frac{\omega-\omega_{0}}{\omega}r_{*}\ ,
\label{eq:hatr}
\end{equation}
\begin{equation}
v=t+\hat{r}
\label{eq:coord_Eddington-Finkelstein}\ ,
\end{equation}
we have the following ingoing wave solution:
\begin{eqnarray}
\Psi_{in} & = & \mbox{e}^{-i \omega v}\mbox{e}^{i \omega \hat{r}}(r-r_{+})^{-\frac{i}{2\kappa_{+}}(\omega-\omega_{0})}\nonumber\\
& = & \mbox{e}^{-i \omega v}\mbox{e}^{i (\omega-\omega_{0}) r_{*}}(r-r_{+})^{-\frac{i}{2\kappa_{+}}(\omega-\omega_{0})}\nonumber\\
& = & \mbox{e}^{-i \omega v}(r-r_{+})^{\frac{i}{2\kappa_{+}}(\omega-\omega_{0})}(r-r_{+})^{-\frac{i}{2\kappa_{+}}(\omega-\omega_{0})}\nonumber\\
& = & \mbox{e}^{-i \omega v}\ .
\label{eq:sol_in_1_Kerr-Newman_tortoise}
\end{eqnarray}
The outgoing wave solution is given by:
\begin{eqnarray}
\Psi_{out}(r>r_{+}) & = & \mbox{e}^{-i \omega v}\mbox{e}^{i \omega \hat{r}}(r-r_{+})^{\frac{i}{2\kappa_{+}}(\omega-\omega_{0})}\nonumber\\
& = & \mbox{e}^{-i \omega v}\mbox{e}^{i (\omega-\omega_{0}) r_{*}}(r-r_{+})^{\frac{i}{2\kappa_{+}}(\omega-\omega_{0})}\nonumber\\
& = & \mbox{e}^{-i \omega v}(r-r_{+})^{\frac{i}{2\kappa_{+}}(\omega-\omega_{0})}(r-r_{+})^{\frac{i}{2\kappa_{+}}(\omega-\omega_{0})}\nonumber\\
& = & \mbox{e}^{-i \omega v}(r-r_{+})^{\frac{i}{\kappa_{+}}(\omega-\omega_{0})}\ .
\label{eq:sol_out_2_Kerr-Newman_tortoise}
\end{eqnarray}

The solutions (\ref{eq:sol_in_1_Kerr-Newman_tortoise}) and (\ref{eq:sol_out_2_Kerr-Newman_tortoise}) are exactly the solutions obtained by Zhang \& Zhao \cite{PhysLettB.618.14} and Huaifan \textit{et al}. \cite{ChinJPhys.47.618}.
%
%
\section{Analytic extension}
In this section, we obtain by analytic continuation a real damped part of the outgoing wave solution of the scalar field which will be used to construct an explicit expression for the decay rate $\Gamma$. This real damped part corresponds (at least in part) to the temporal contribution to the decay rate \cite{PhysLettB.666.269,IntJModPhysD.17.2453} found in the tunneling method of Hawking radiation.

From Eq.~(\ref{eq:sol_out_2_Kerr-Newman_tortoise}), we see that this solution is not analytical in the exterior event horizon $r=r_{+}$. By analytic continuation, rotating $-\pi$ through the lower-half complex $r$ plane, we obtain
\begin{equation}
(r-r_{+}) \rightarrow \left|r-r_{+}\right|\mbox{e}^{-i\pi}=(r_{+}-r)\mbox{e}^{-i\pi}\ .
\label{eq:rel_3_Kerr-Newman}
\end{equation}
Thus, the outgoing wave solution on the horizon surface $r_{+}$ is
\begin{equation}
\Psi_{out}(r<r_{+})=\mbox{e}^{-i\omega v}(r_{+}-r)^{\frac{i}{\kappa_{+}}(\omega-\omega_{0})}\mbox{e}^{\frac{\pi}{\kappa_{+}}(\omega-\omega_{0})}\ .
\label{eq:sol_1_out_4_Kerr-Newman}
\end{equation}
Now, using Eq.~(\ref{eq:coord_tortoise_1}), the solution given by Eq.~(\ref{eq:sol_1_out_4_Kerr-Newman}) can also be written in the following form
\begin{equation}
\Psi_{out}(r<r_{+})=\mbox{e}^{-i\omega v}\mbox{e}^{2i(\omega-\omega_{0})r_{*}}\mbox{e}^{\frac{\pi}{\kappa_{+}}(\omega-\omega_{0})}\ .
\label{eq:sol_2_out_4_Kerr-Newman}
\end{equation}
Eqs. (\ref{eq:sol_out_2_Kerr-Newman_tortoise}) and (\ref{eq:sol_2_out_4_Kerr-Newman}) describe the outgoing wave outside and inside of the black hole, respectively. Therefore, for an outgoing wave of a particle with energy $\omega$, charge $e$ and angular momentum $m$, the outgoing decay rate or the relative scattering probability of the scalar wave at the event horizon surface $r=r_{+}$ is given by
\begin{equation}
\Gamma_{+}=\left|\frac{\Psi_{out}(r>r_{+})}{\Psi_{out}(r<r_{+})}\right|^{2}=\mbox{e}^{-\frac{2\pi}{\kappa_{+}}(\omega-\omega_{0})}\ .
\label{eq:taxa_refl_Kerr-Newman}
\end{equation}

This result was formally obtained in the literature \cite{PhysLettB.618.14,ChinJPhys.47.618,PhysLettB.666.269,IntJModPhysD.17.2453} in different contexts.
%
%
\section{Radiation spectrum}
After the black hole event horizon radiates particles with energy $\omega$, charge $e$ and angular momentum $m$, in order to consider the reaction of the radiation of the particle to the spacetime, we must replace $M,Q,J$ by $M-\omega,Q-e,J-m$, respectively, in the line element of Kerr-Newman spacetime (\ref{eq:metrica_Kerr-Newman}). Doing these changes, we must guarantee that the total energy, angular momentum and charge of spacetime are all conserved, that is,
\begin{equation}
\begin{array}{r}
-\omega=\Delta E\ ,\\
-e=\Delta Q\ ,\\
-m=\Delta J\ ,
\end{array}
\label{eq:param_cons_Kerr-Newman}
\end{equation}
where $\Delta E$, $\Delta Q$, and $\Delta J$ are the energy, charge, and angular momentum variations of the black hole event horizon, before and after the emission of radiation, respectively. Substituting Eqs.~(\ref{eq:1_lei_termo_Kerr-Newman}) and (\ref{eq:param_cons_Kerr-Newman}) into Eq.(\ref{eq:taxa_refl_Kerr-Newman}), we obtain the outgoing decay rate at the event horizon surface $r=r_{+}$:
\begin{eqnarray}
\Gamma_{+} & = & \mbox{e}^{-\frac{2\pi}{\kappa_{+}}(-\Delta E-m\Omega_{+}-e\Phi_{+})}\nonumber\\
& = & \mbox{e}^{-\frac{2\pi}{\kappa_{+}}(-\Delta E+\Omega_{+}\Delta J+\Phi_{+}\Delta Q)}\nonumber\\
& = & \mbox{e}^{-\frac{2\pi}{\kappa_{+}}(-T_{+}\Delta S_{+})}\nonumber\\
& = & \mbox{e}^{\Delta S_{+}}\ ,
\label{eq:taxa_refl_param_Kerr-Newman}
\end{eqnarray}
where we have used Eq.~(\ref{eq:temp_Hawking_Kerr-Newman}). $\Delta S_{+}$ is the change of the Bekenstein-Hawking entropy, compared before and after the emission of radiation, and obtained from the expressions for the entropy (\ref{eq:entropia_Kerr-Newman}) and for the exterior event horizon (\ref{eq:sol_padrao_Kerr-Newman_1}), as follows:
\begin{eqnarray}
\Delta S_{+} & = & S_{+}(M-\omega,Q-e,J-m)-S_{+}(M,Q,J)\nonumber\\
& = & \pi\left[r_{+}^{2}(M-\omega,Q-e,J-m)+a^{2}(M-\omega,J-m)\right]\nonumber\\
& - & \pi\left[r_{+}^{2}(M,Q,J)+a^{2}(M,J)\right]\nonumber\\
& = & \pi[2(M-\omega)^{2}-(Q-e)^{2}\nonumber\\
& + & 2(M-\omega)\sqrt{(M-\omega)^{2}-a_{\omega}^{2}-(Q-e)^{2}}\nonumber\\
& + & Q^{2}-2M^{2}-2M\sqrt{M^{2}-a^{2}-Q^{2}}]\ ,
\label{eq:entropia_Bekenstein-Hawking_Kerr-Newman}
\end{eqnarray}
where $a=a(M,J)=J/M$ and
\begin{equation}
a_{\omega}^{2}=\left(\frac{J-m}{M-\omega}\right)^{2}\ .
\label{eq:a_omega_Kerr-Newman}
\end{equation}
According to the Damour-Ruffini-Sannan method \cite{PhysRevD.14.332,GenRelativGravit.20.239}, a correct wave describing a particle flying off of the black hole is given by
\begin{eqnarray}
\Psi_{\omega}(r) & = & N_{\omega}\ [\ H(r-r_{+})\ \Psi_{\omega}^{out}(r-r_{+})\nonumber\\
& + & H(r_{+}-r)\ \Psi_{\omega}^{out}(r_{+}-r)\ \mbox{e}^{\frac{\pi}{\kappa_{+}}(\omega-\omega_{0})}\ ]\ ,
\label{eq:solucao_geral_onda_out_Kerr-Newman}
\end{eqnarray}
where $N_{\omega}$ is the normalization constant, such that
\begin{equation}
\left\langle \Psi_{\omega_{1}}(r) | \Psi_{\omega_{2}}(r) \right\rangle=-\delta(\omega_{1}-\omega_{2})\ ,
\label{eq:cond_norm}
\end{equation}
where $H(x)$ is the Heaviside function and $\Psi_{\omega}^{out}(x)$ are the normalized wave functions given, from Eq.~(\ref{eq:sol_out_2_Kerr-Newman_tortoise}), by
\begin{equation}
\Psi_{\omega}^{out}(x)=\mbox{e}^{-i \omega v}x^{\frac{i}{\kappa_{+}}(\omega-\omega_{0})}\ .
\label{eq:sol_out_2_Kerr-Newman_tortoise_x}
\end{equation}

Thus, from the normalization condition
\begin{equation}
\left\langle \Psi_{\omega}(r) | \Psi_{\omega}(r) \right\rangle=1=\left|N_{\omega}\right|^{2}\left[\mbox{e}^{\frac{2\pi}{\kappa_{+}}(\omega-\omega_{0})}-1\right]\ ,
\label{eq:norm_onda_out_Kerr-Newman}
\end{equation}
we get the resulting Hawking radiation spectrum of charged scalar particles
\begin{equation}
\left|N_{\omega}\right|^{2}=\frac{1}{\mbox{e}^{\frac{2\pi}{\kappa_{+}}(\omega-\omega_{0})}-1}=\frac{1}{\mbox{e}^{\frac{\hbar(\omega-\omega_{0})}{k_{B}T_{+}}}-1}\ .
\label{eq:espectro_rad_Kerr-Newman_2}
\end{equation}
Therefore, we can see that the resulting Hawking radiation spectrum of scalar particles has a thermal character, analogous to the blackbody spectrum, where $k_{B}T_{+}=\hbar\kappa_{+}/2\pi$, with $k_{B}$ being the Boltzmann constant. Note that Planck and Boltzmann's constants were introduced into Eq.~(\ref{eq:espectro_rad_Kerr-Newman_2}) in order to have the correct dimension.
%
%
\section{Conclusions}
In this paper, we presented analytic solutions for both angular and radial parts of the Klein-Gordon equation for a charged massive scalar field in the Kerr-Newman spacetime.

These solutions are analytic in whole spacetime, namely, in the region between the event horizon and infinity. The radial part of the obtained solutions generalizes a previous result \cite{ClassQuantumGrav.31.045003} in the sense that now we are considering a charged massive scalar field coupled to the electromagnetic field associated with the gravitational source. The solution is given in terms of the confluent Heun functions, and is valid over the range $0 \leq x < \infty$.

From these analytic solutions, we obtained the solutions for ingoing and outgoing waves near the exterior horizon of a Kerr-Newman black hole, and used these results to discuss the Hawking radiation effect \cite{CommunMathPhys.43.199}. We considered the properties of the confluent Heun functions to obtain the results. This approach has the advantage that it is not necessary the introduction of any coordinate system, as for example, the particular one \cite{PhysLettB.618.14} or tortoise and Eddington-Finkelstein coordinates \cite{ChinJPhys.47.618}.
%
%
\section*{Acknowledgement}
The authors would like to thank Conselho Nacional de Desenvolvimento Cient\'{i}fico e Tecnol\'{o}gico (CNPq) for partial financial support.
%
%

%
%

\begin{thebibliography}{99}
\bibitem{AnnPhys.386.109} E. Schr\"{o}dinger, Ann. Phys. (Berlin) \textbf{386}, 109 (1926)
\bibitem{ZPhys.37.895} O. Klein, Z. Phys. \textbf{37}, 895 (1926)
\bibitem{ZPhys.40.117} W. Gordon, Z. Phys. \textbf{40}, 117 (1926)
\bibitem{ZPhys.41.407} O. Klein, Z. Phys. \textbf{41}, 407 (1927)
\bibitem{ProcRSocLondA.117.610} P. A. M. Dirac, Proc. R. Soc. Lond. A \textbf{117}, 610 (1928)
\bibitem{ProcRSocLondA.118.351} P. A. M. Dirac, Proc. R. Soc. Lond. A \textbf{118}, 351 (1928)
\bibitem{ZPhys.53.592} E. Wigner, Z. Phys. \textbf{53}, 592 (1929)
\bibitem{ZPhys.57.261} V. Fock, Z. Phys. \textbf{57}, 261 (1929)
\bibitem{AnnPhys.410.305} W. Pauli, Ann. Phys. (Berlin) \textbf{410}, 305 (1933)
\bibitem{AnnPhys.410.337} W. Pauli, Ann. Phys. (Berlin) \textbf{410}, 337 (1933)
\bibitem{Physica.6.899} E. Schr\"odinger, Physica \textbf{6}, 899 (1939)
%
\bibitem{rowan2} D. J. Rowan and G. Stephenson, J. Phys. A: Math. Gen. \textbf{9}, 1261 (1976)
\bibitem{stephenson} D. J. Rowan and G. Stephenson, J. Phys. A: Math. Gen. \textbf{9}, 1631 (1976)
\bibitem{JPhysAMathGen.10.15} D. J. Rowan and G. Stephenson, J. Phys. A: Math. Gen. \textbf{10}, 15 (1977)
\bibitem{JMathPhys.22.1457} W. E. Couch, J. Math. Phys. \textbf{22}, 1457 (1981)
\bibitem{JMathPhys.26.2286} W. E. Couch, J. Math. Phys. \textbf{26}, 2286 (1985)
\bibitem{pimentel} L. O. Pimentel and A. Mac\'{i}as, Phys. Lett. A \textbf{117}, 325 (1986)
\bibitem{elizalde} E. Elizalde, Phys. Rev. D \textbf{36}, 1269 (1987)
\bibitem{pimentel2} L. O. Pimentel, Gen. Relativ. Grav. \textbf{24}, 985 (1992)
\bibitem{semiz} I. Semiz, Phys. Rev. D. \textbf{45}, 532 (1992)
\bibitem{kraori} K. D. Kraori, P. Borgohain and D. Das, J. Math. Phys. \textbf{35}, 1032 (1994)
\bibitem{JMathPhys.40.4538} S. Q. Wu and X. Cai, J. Math. Phys. \textbf{40}, 4538 (1999)
\bibitem{ProgTheorPhys.112.983} H. Furuhashi and Y. Nambu, Prog. Theor. Phys. \textbf{112}, 983 (2004)
\bibitem{bezerra} R. M. Teixeira Filho and V. B. Bezerra, Class. Quantum Grav. \textbf{21}, 307 (2004)
\bibitem{vakili} A. Alimohammadi and B. Vakili, Ann. Phys. (NY) \textbf{310}, 95 (2004)
\bibitem{sandro} S. G. Fernandes, G. A. Marques and V. B. Bezerra, Class. Quantum Grav. \textbf{23}, 7063 (2006)
\bibitem{GenRelativGrav.43.833} I. Semiz, Gen. Relativ. Grav. \textbf{43}, 833 (2011)
%
\bibitem{ClassQuantumGrav.31.045003} V. B. Bezerra, H. S. Vieira and Andr\'{e} A. Costa, Class. Quantum Grav. \textbf{31}, 045003 (2014)
\bibitem{Ronveaux:1995} A. Ronveaux, \textit{Heun's differential equations}, (Oxford University Press, New York, 1995)
\bibitem{Slavyanov:2000} S. Y. Slavyanov and W. Lay, \textit{Special functions, A unified theory based on singularities}, (Oxford University Press, New York, 2000)
\bibitem{CommunMathPhys.43.199} S. W. Hawking, Commun. Math. Phys. \textbf{43}, 199 (1975)
%
\bibitem{PhysRevD.11.1404} D. G. Boulware, Phys. Rev. D \textbf{11}, 1404 (1975)
\bibitem{PhysRevD.13.2188} J. B. Hartle and S. W. Hawking, Phys. Rev. D \textbf{13}, 2188 (1976)
\bibitem{PhysRevD.14.332} T. Damour and R. Ruffini, Phys. Rev. D \textbf{14}, 332 (1976)
\bibitem{PhysRevD.15.2738} G. W. Gibbons and S. W. Hawking, Phys. Rev. D \textbf{15}, 2738 (1977)
\bibitem{ProcRSocLondA.15.2738} G. W. Gibbons and M. J. Perry, Proc. R. Soc. Lond. A \textbf{358}, 467 (1978)
\bibitem{GenRelativGravit.20.239} S. Sannan, Gen. Relativ. Gravit. \textbf{20}, 239 (1988)
\bibitem{PhysRevLett.85.5042} M. K. Parikh and F. Wilczek, Phys. Rev. Lett. \textbf{85}, 5042 (2000)
\bibitem{PhysLettB.642.124} E. T. Akhmedov, V. Akhmedova and D. Singleton, Phys. Lett. B \textbf{642}, 124 (2006)
\bibitem{IntJModPhysA.22.1705} E. T. Akhmedov, V. Akhmedova, D. Singleton and T. Pilling, Int. J. Mod. Phys. A. \textbf{22}, 1705 (2007)
%
\bibitem{PhysRevD.82.044013} R. Banerjee, C. Kiefer and B. R. Majhi, Phys. Rev. D \textbf{82}, 044013 (2010)
\bibitem{PhysLettB.692.61} K. Umetsu, Phys. Lett. B \textbf{692}, 61 (2010)
\bibitem{PhysLettB.697.398} A. Yale, Phys. Lett. B \textbf{697}, 398 (2011)
%
\bibitem{PhysLettB.618.14} J. Y. Zhang and Z. Zhao, Phys. Lett. B \textbf{618}, 14 (2005)
\bibitem{ChinJPhys.47.618} H. F. Li, S. L. Zhang, and R. Zhao, Chin. J. Phys. \textbf{47}, 618 (2009)
\bibitem{IntJModPhysA.25.4123} K. Umetsu, Int. J. Mod. Phys. A. \textbf{25}, 4123 (2010)
%
\bibitem{ProgTheorPhys.100.491} H. Suzuki, E. Takasugi and H. Umetsu, Prog. Theor. Phys. \textbf{100}, 491 (1998)
\bibitem{ProgTheorPhys.102.253} H. Suzuki, E. Takasugi and H. Umetsu, Prog. Theor. Phys. \textbf{102}, 253 (1999)
\bibitem{JMathPhys.48.042502} D. Batic and H. Schmid, J. Math. Phys. \textbf{48}, 042502 (2007)
\bibitem{ClassQuantumGrav.27.135001} P. P. Fiziev, Class. Quantum Grav. \textbf{27}, 135001 (2010)
%
\bibitem{AstrophysicalJ.185.635} S. A. Teukolsky, Astrophysical J. \textbf{185}, 635 (1973)
%
\bibitem{MTW:1973} C. W. Misner, K. S. Thorne and J. A. Wheeler, \textit{Gravitation}, (W. H. Freeman and Company, San Francisco, 1973)
\bibitem{JMathPhys.8.265} R. H. Boyer and R. W. Lindquist, J. Math. Phys. \textbf{8}, 265 (1967)
\bibitem{ChinAstronAstrophys.5.365} Z. Zhao and Y. X. Guei, Chin. Astron. Astrophys. \textbf{5}, 365 (1981)
\bibitem{PhysRevD.12.2963} L. H. Ford, J. Math. Phys. \textbf{12}, 2963 (1975)
\bibitem{Morse:1953} P. M. Morse and H. Feschbach, \textit{Methods of Theorical Physics, Part II}, (McGraw Hill, New York, 1953)
\bibitem{JPhysAMathTheor.43.035203} P. P. Fiziev, J. Phys. A: Math. Theor. \textbf{43}, 035203 (2010)
\bibitem{PhysLettB.666.269} V. Akhmedova, T. Pilling, A. Gill and D. Singleton, Phys. Lett. B \textbf{666}, 269 (2008)
\bibitem{IntJModPhysD.17.2453} E. T. Akhmedov, T. Pilling and D. Singleton, Int. J. Mod. Phys. D \textbf{17}, 2453 (2008)
\end{thebibliography}
\end{document}